\documentclass{iau}

\usepackage{amsmath}
\usepackage{graphicx}
\usepackage{multirow}
\usepackage{xcolor}
\usepackage{subfigure}
\newcommand{\HDtwentyone}{HD\,215575}

\newcommand{\HDtwentyfive}{HD\,253659}

\newcommand{\BD}{BD$+$14\,1106}
\newcommand{\bd}{BD$+$14\,1106}

\usepackage{orcidlink}
\begin{document}

\lefttitle{Cambridge Author}
\righttitle{Proceedings of the International Astronomical Union: \LaTeX\ Guidelines for~authors}

\jnlPage{1}{7}
\jnlDoiYr{2021}
\doival{10.1017/xxxxx}

\volno{402}

\aopheadtitle{Proceedings IAU Symposium}
\editors{A. Wofford,  N. St-Louis, M. Garcia \&  S. Simón-Díaz, eds.}

\title{Newly discovered nebulae around Galactic B-type stars and their origins}

\author{Olga~Maryeva\orcidlink{0000-0003-1442-4755}$^{1}$, Julieta~S\'{a}nchez Arias\orcidlink{0000-0003-3599-516X}$^{1}$, Sergey~Karpov\orcidlink{0000-0003-0035-651X}$^{2}$, Michaela~Kraus\orcidlink{0000-0002-4502-6330}$^{1}$, Sabina Mammadova$^{3}$,   Michalis~Kourniotis$^{1}$,  P\'eter N\'emeth\orcidlink{0000-0003-0963-0239}$^{4}$, Lydia~Cidale\orcidlink{0000-0003-2160-7146}$^{5}$,  Anahi~Granada$^{6,7}$     }

\affiliation{1 Astronomical Institute of the Czech Academy of Sciences, Fri\v{c}ova 298, 25165 Ond\v{r}ejov, Czech Republic \\
2 Institute of Physics of the  Czech Academy of Sciences, CZ-182 21 Prague 8, Czech Republic \\
3 Shamakhy Astrophysical Observatory, Y.Mammadaliyev, AZ5626, Azerbaijan \\
4 \quad {Astroserver.org}, F\H{o}t\'er 1, 8533 Malomsok, Hungary \\
5 Instituto de Astrof\'{i}sica de La Plata, Paseo del Bosque S/N, La Plata, Argentina \\
%6 Laboratorio de Procesamiento de Se\~{n}ales Aplicadas y Computaci\'{o}n de Alto Rendimiento, Sede Andina, Universidad Nacional de R\'{i}o Negro, San Carlos de Bariloche, Argentina}
6 Centro Interdisciplinario de Telecomunicaciones, Electr\'{o}nica, Computaci\'{o}n y Ciencia Aplicada - Sede Andina, Universidad Nacional de R\'{i}o Negro, San Carlos de Bariloche, Argentina. \\
7  Consejo Nacional de Investigaciones Científicas y T\'{e}cnicas, Ciudad Autónoma de Buenos Aires, Argentina. }

\begin{abstract}
The mass loss in massive stars is an important process that determines their future evolution and affects their circumstellar environments. Besides the continuous outflow of matter in the form of stellar winds, massive stars undergo sporadic mass ejections that lead to the formation of circumstellar envelopes. For now it remains unclear at what stage of evolution the first mass ejection occurs and what instabilities lead to it. We present the results of a study of four B-type stars with  circumstellar nebulae that have recently been found in the archive of the Wide-field Infrared Survey Explorer (WISE), as well as  a list of newly detected nebulae around classical Be stars and $\beta$\,Cephei variables up to 13 mag.
%
%{\color{red} Two of our objects -- PY\,Gem and HD\,253659 -- are eruptive $\gamma$\,Cas variables showing double peaked emission H$\alpha$ profiles, and during our analysis we found that such nebulae are typical for $\gamma$\,Cas variables. The other two stars -- HD\,215575 and BD+14\,1106 -- have spectra of normal B-type stars on the main sequence. Spectral analysis, numerical modeling (with Tlusty code), as well as their high proper motions argue that these two objects undergone merging in the past.}
\end{abstract}


\maketitle

\section{Introduction}

To date, the all-sky infrared (IR) survey performed by the {\it WISE} space telescope is the main source for detecting new dusty nebulae. IR circumstellar nebulae, in combination with the blue colour of the central star, are widely used to identify evolved massive stars, including merger products  {\citep[for details see][]{GKF2010, Gvaramadze2017, Gvaramadze2019Nature}.

In this work, we consider four stars with newly detected circumstellar nebulae. Table~\ref{tab:stars} lists the parameters of the stars and their spectral classification taken from the literature, while Fig.~\ref{mmt} presents their light curves.  Two of our objects -- PY\,Gem and HD\,253659\ -- are classical Be stars showing double-peaked emission  profiles of H$\alpha$ and eruptive photometric activity ($\gamma$\,Cas variables). 
The bump in HD\,253659 light curve between 2016 and 2022 may be interpreted as a major outburst. The spectrum of HD\,215575 corresponds to B1\,I type but the apparent magnitude does not match the expected luminosity of a supergiant. Therefore we suggest that HD\,215575 is a case of spectral mimicry, being a runaway low-mass post-AGB star with a spectral mimicry to a B1 supergiant. The TESS light curve of \bd\ displays a regular short period, which can be interpreted as intrinsic pulsations typical of $\beta$\,Cephei (BCEP) stars. But in addition, it contains a periodicity which may be the evidence of binarity. This binarity might also explain the origin and bipolar shape of its nebula. 

These newly discovered nebulae (upper panel of Fig.~\ref{fig:nebulae}) motivated us to conduct a review of known galactic Be stars, $\gamma$\,Cas variables (i.e. eruptive Be) and BCEPs. We analysed {\it WISE} IR images of 1438 Be stars, 631 $\gamma$\,Cas variables and 1405 BCEPs. As a result, we found 52 / 44 / 37 circumstellar nebulae around Be stars, $\gamma$\,Cas variables, BCEPs respectively, and 46 / 27 / 22 of them were detected for the first time. 
The full list of stars with detected circumstellar nebulae, including their images, is available at {\url{https://stel.asu.cas.cz/~maryeva/Nebulae/}. 

A statistical analysis of the nebulae around Be and $\gamma$\,Cas variables shows that the nebulae are characteristic for them, although mostly visible for brighter stars only. The stars with and without nebulae share the same location in the HR diagram, close to the main sequence. 
The fraction of binary stars among stars with nebulae is the same as in the general Be star population. We compared the multiplicity according the {\it Gaia} and {\it Hipparcos} in the Be samples with and without nebulae, and found in both the same fraction of $21\%\pm7.5$, which is in good agreement with the multiplicity of Be stars \citep{Dodd2024, Guerrero2025}.
%Dodd et al.2024
Thus, the nebulae are probably inherent to the Be phenomenon and caused by the interplay between their rapid rotation and pulsations.

%-----------------------------------
\begin{table*}\centering
\small
\caption{Stars studied in this work.  Dist is the distance to the objects according to \citet{Bailer-Jones2021}, $\mu$ is the proper motions from \citet{GaiaEDR3}. \label{tab:stars} }
%\resizebox{\textwidth}{!}
{\begin{tabular}{l lc ccc cc }
%\toprule
\hline
\multicolumn{1}{c}{\multirow{2}{*}{Star}} 
&  \multicolumn{1}{c}{\multirow{2}{*}{Sp. type}} & Var.& Dist & $\mu$          & \multicolumn{1}{c}{\multirow{2}{*}{RUWE}}    &  $V$& Size \\
    &                                               & type& [pc] & [mas yr$^{-1}$]&  &  [mag] & [pc] \\ 
\hline
PY\,Gem         & B1~Vne$^{a}$      & $\gamma$\,Cas & $1401_{-63}^{+48}$  & 1.301 & 0.977 &8.46 & 1.83\\ % Gaia DR3 
\BD             &  B2~V$^{b}$       & -- & $1693_{-83}^{+99}$  &2.710  & 1.018  & 9.62 & 1.23\\%  APASS
\HDtwentyone    & B8$^{c}$            & -- & $754\pm10$& 29.919 & 0.906 & 8.05 & 1.5 \\ %Gaia DR3
\HDtwentyfive   & B0.5~V:nne$^{d}$  & $\gamma$\,Cas  & $1698_{-58}^{+66}$  & 1.002 & 1.106 & 9.16  & 1.48 \\ %Gaia DR3 
\hline
% \multicolumn{9}{l}{$^a$ \citet{Guetter1968}; $^b$ \citet{SpectralClassification2020}; $^c$ \citet{CannonPickering1993}; $^d$ \citet{Morgan1955}. }\\
\end{tabular}}
{\begin{tabular}{l}
{The photometry is taken from the Gaia DR3 Synthetic Photometry catalogue  \citep{gaiadr3syn},}\\
{ except for \bd\ where the photometry is from APASS DR10 \citep{vo:apass_dr10}. Variability type } \\ 
{is according to AAVSO Variable Stars Index \citep{VariableStarVSX2006}. Spectral types are  from: $^{a}$ {\citet{Guetter1968}},} \\
{$^b$ \citet{SpectralClassification2020}, 
$^c$ \citet{CannonPickering1993}, $^{d}$ {\citet{Morgan1955}}. }
\end{tabular}}
\end{table*}
\vspace{-0.05cm}
%-----------------------------------

%We selected these objects {\color{orange} for separate study} based on their brightnesses ($V<10$~mag), which allows their study using medium and high resolution spectroscopy.  For 

%%%%%%%%%%%%%%%%%%%%%%%%%%%%%%%%
  \begin{figure*}
   \centering
   %\centerline{\resizebox*{1.0\columnwidth}{!}{\includegraphics[angle=0]{ztf_curve.eps}}}
   \includegraphics[width=1.0\columnwidth,clip]{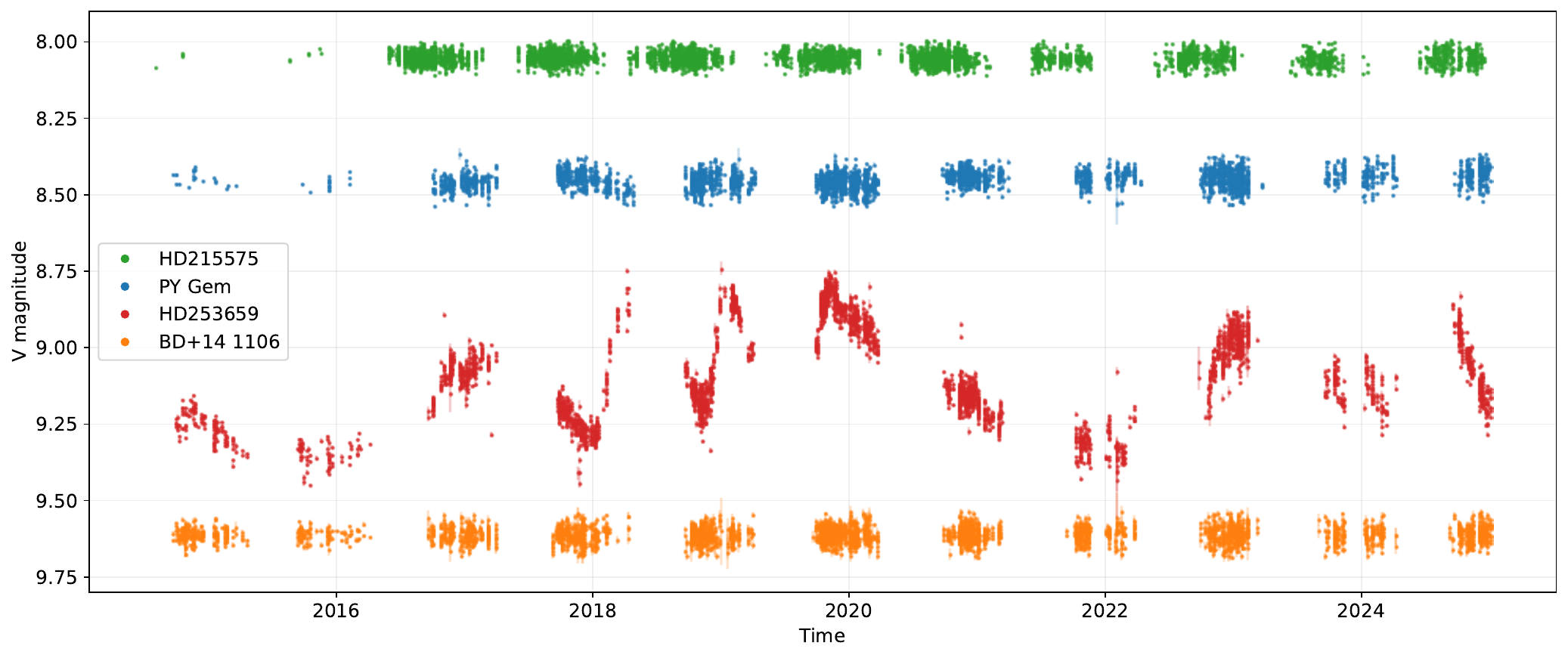}
   \vspace{-0.7cm}
   \caption{Light curves of the studied stars from the Mini-MegaTORTORA \citep{MMT2017} archive ({\url{http://survey.favor2.info/}}). \label{mmt}}
   \end{figure*}
    \vspace{-0.3cm}
%%%%%%%%%%%%%%%%%%%%%%%%%%%%%%%%%%%%%%%%%%%%%%%%

%%%%%%%%%%%%%%%%%%%%%%%%%% FIGURE %%%%%%%%%%%%%%%%%%%%%%%%%%%%%
\begin{figure}
\vspace{-0.2cm}
%\resizebox*{0.65\columnwidth}{!}{\includegraphics[angle=0,viewport=15 25 750 750,clip]{BD_5.eps}}
\subfigure[PY\,Gem ($\gamma$\,Cas)]{\resizebox*{0.245\columnwidth}{!}{\includegraphics{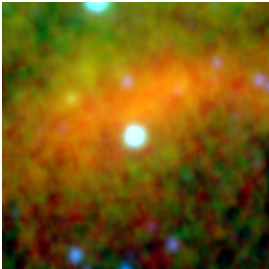}}}
\subfigure[HD\,253659 ($\gamma$\,Cas)]{\resizebox*{0.245\columnwidth}{!}{\includegraphics{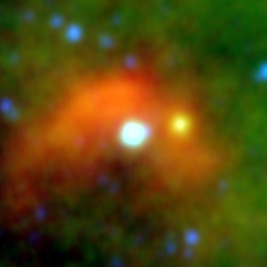}}} 
\subfigure[{$\rm \bd$} (BCEP)]{\resizebox*{0.245\columnwidth}{!}{\includegraphics{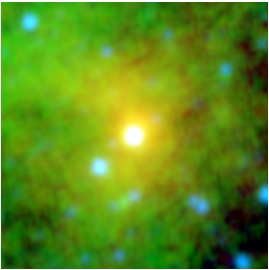}}}
\subfigure[{HD\,215575}]{\resizebox*{0.245\columnwidth}{!}{\includegraphics{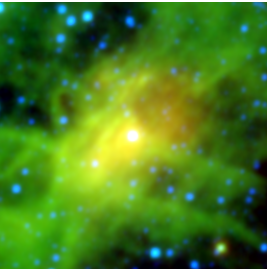}}}
\vspace{-0.42cm}\\
%\subfigure[HD\,27846]{\resizebox*{0.245\columnwidth}{!}{\includegraphics{HD_27846_rgb_1.png}}} 
%\subfigure[HD\,331976]{\resizebox*{0.245\columnwidth}{!}{\includegraphics{HD_331976_rgb_1.png}}}
\subfigure[$\rm \epsilon$\,Per (BCEP)]{\resizebox*{0.245\columnwidth}{!}{\includegraphics{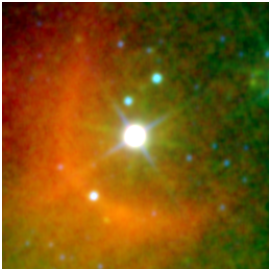}}}
\subfigure[HD\,57682 (Be)]{\resizebox*{0.245\columnwidth}{!}{\includegraphics{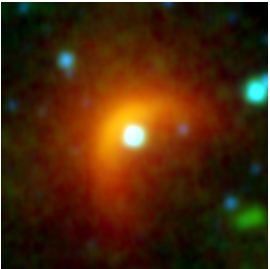}}}
\subfigure[HD\,84567 (Be)]{\resizebox*{0.245\columnwidth}{!}{\includegraphics{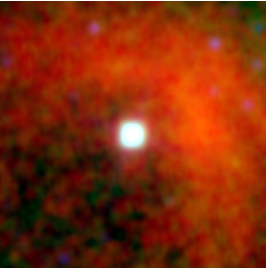}}}
\subfigure[FN\,CMa (BCEP)]{\resizebox*{0.245\columnwidth}{!}{\includegraphics{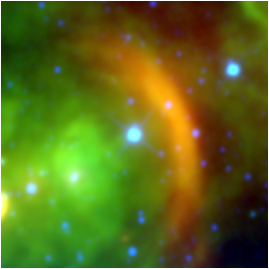}}}
%\subfigure[$\sigma$\,Sco]{\resizebox*{0.245\columnwidth}{!}{\includegraphics{sig_Sco_rgb_1.png}}}
\vspace{-0.4cm}\\
\subfigure[$\rm \delta$\,Sco ($\gamma$\,Cas)]{\resizebox*{0.245\columnwidth}{!}{\includegraphics{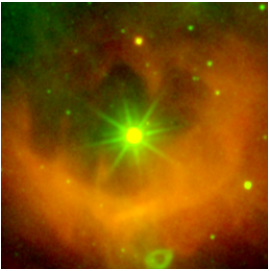}}} %!!!
\subfigure[V421\,Cep (Be)]{\resizebox*{0.245\columnwidth}{!}{\includegraphics{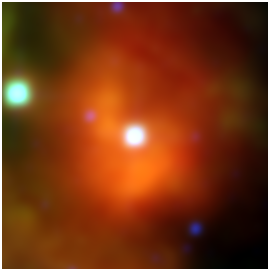}}}
\subfigure[V750\,Mon ($\gamma$\,Cas)]{\resizebox*{0.245\columnwidth}{!}{\includegraphics{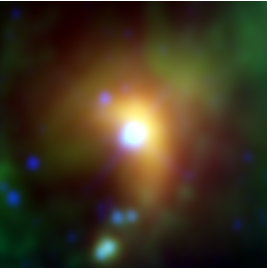}}} % new
\subfigure[V4375\,Sgr (Be)]{\resizebox*{0.245\columnwidth}{!}{\includegraphics{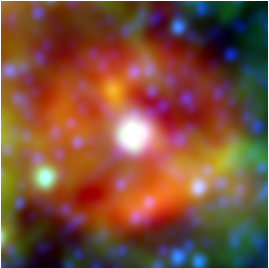}}}
\vspace{-0.4cm}
\caption{Examples of RGB images with WISE 22-$\mu$m (red channel), WISE 12-$\mu$m (green channel), WISE 4.6-$\mu$m (blue channel) bands showing structures the discovered nebulae.} \label{fig:nebulae}
\end{figure}   
%%%%%%%%%%%%%%%%%%%%%%%%%%%%%%%%%%%%%%%%%%%%%%%%%%%%%%%%

%%%%%%%%%%%%%%%%%%%%%%%%%%%%%%%%%%%%%%%%%%%%%%%%%%%%%%%%%%
\section*{Acknowledgements}
%%%%%%%%%%%%%%%%%%%%%%%%%%%%%%%%%%%%%%%%%%%%%%%%%%%%%%%%%%
We acknowledge support from the Czech Science Foundation  GA\v{C}R 25-17532S. The Astronomical Institute in Ond\v{r}ejov is supported by the project RVO:67985815. 
S.K. acknowledges support by the European Union and the Czech Ministry of Education, Youth and Sports (Project No. CZ.02.01.01/00/22\_008/0004632 -- FORTE).
This project is co-funded by the European Union (Project 101183150 - OCEANS). 

\bibliographystyle{aa} % style aa.bst
\bibliography{Nebulae} % your references Yourfile.bib

\end{document}